\documentclass[aps,prl,preprintnumbers,nofootinbib,superscriptaddress,twocolumn]{revtex4-1}
\usepackage{natbib}
\usepackage{graphicx}
\usepackage{xcolor}

\usepackage[flushleft]{threeparttable}

\def\prd{Phys. Rev. D}
\def\mnras{MNRAS}
\def\apj{ApJ}
\def\apjl{ApJL}
\def\apjs{ApJS}
\def\aap{A\&A}

\def\jcap{JCAP}
\def\pasa{PASA}

\begin{document}

\preprint {HRI-RECAPP-2018-014}

\title{Constraints on dark matter annihilation in dwarf spheroidal galaxies from low frequency radio observations}

\author{Arpan Kar}
\email{arpankar@hri.res.in}
\affiliation{Regional Centre for Accelerator-based Particle Physics, Harish-Chandra Research Institute, HBNI, Chhatnag Road, Jhunsi, Allahabad - 211 019, India}

\author{Sourav Mitra}
\affiliation{Surendranath College, 24/2 M. G. ROAD, Kolkata, West Bengal 700009, India}

\author{Biswarup Mukhopadhyaya}
\affiliation{Regional Centre for Accelerator-based Particle Physics, Harish-Chandra Research Institute, HBNI, Chhatnag Road, Jhunsi, Allahabad - 211 019, India}

\author{Tirthankar Roy Choudhury}
\affiliation{National Centre for Radio Astrophysics, TIFR, Post Bag 3, Ganeshkhind, Pune 411007, India}

\author{Steven Tingay}
\affiliation{International Centre for Radio Astronomy Research, Curtin University, Bentley, WA 6102, Australia}

\begin{abstract}
We present the first observational limits on the predicted synchrotron signals from particle Dark Matter annihilation models in dwarf spheroidal galaxies at radio frequencies below 1 GHz.  We use a combination of survey data from the Murchison Widefield Array (MWA) and the Giant Metre-wave Radio Telescope (GMRT) to search for diffuse radio emission from 14 dwarf spheroidal galaxies.  For in-situ magnetic fields of 1 $\mu G$ and any plausible value for the diffusion coefficient, our limits do not constrain any Dark Matter models. However, for stronger magnetic fields our data might provide constraints comparable to existing limits from gamma-ray and cosmic ray observations. Predictions for the sensitivity of the upgraded MWA show that models with Dark Matter particle mass up to $\sim$ 1.6 TeV (1 TeV) may be constrained for magnetic field of 2 $\mu G$ (1 $\mu G$). While much deeper limits from the future low frequency Square Kilometre Array (SKA) will challenge the LHC in searches for Dark Matter particles, the MWA provides a valuable first step toward the SKA at low frequencies.
\end{abstract}


\maketitle


{\it Introduction: The synchrotron radiation produced because of self-annihilating dark matter (DM) candidate particles in dwarf spheroidal (dSph) galaxies (objects with high mass-to-light ratios indicating a high abundance of DM) can be a promising probe of DM models. \cite{Kar:2019mcq} explore the use of the Square Kilometre Array (SKA) for the detection of synchrotron signatures from dSphs (Draco, Segue I, and Ursa Major II); they demonstrate that the SKA could significantly exceed the reach of the Large Hadron Collider (LHC) in the search for self-annihilating DM candidate particles that produce charged particles and hence synchrotron emission due to an in-situ magnetic field. Such predicted synchrotron signals were discussed earlier by \cite{2015aska.confE.100C}, but for masses within the LHC reach. In this context, we analyse here some data from the Murchison Widefield Array (MWA), a precursor to the SKA.} 

Using Green Bank Telescope (GBT) observations at 1.4 GHz, \cite{2015arXiv150703589N} derive upper limits on radio synchrotron emission from Segue I and conclude that annihilation to $e^{+}e^{-}$ is strongly disfavoured for DM particle masses $<50$ GeV, but that other annihilation channels are not strongly constrained.  \cite{2017JCAP...07..025R,2014JCAP...10..016R} used the Australia Telescope Compact Array to search for similar signals from nearby dSph galaxies accessible from the Southern Hemisphere.

These observations were conducted at trans-GHz frequencies, whereas the synchrotron signal is expected to be stronger at lower frequencies \cite{Kar:2019mcq}.  However, robust attempts to measure the DM annihilation synchrotron signal at low frequencies are currently lacking in the literature.  The work described in this paper addresses this deficiency for the first time, heralded by the emergence of modern low frequency facilities such as the MWA, a precursor for the even larger future SKA. The synchrotron signal expected to accompany DM annihilation is diffuse in nature, following the DM distribution. Therefore, high surface brightness radio observations are required. The observations with the GBT, a single dish, have excellent surface brightness sensitivity.  The observations with the ATCA, as a relatively sparse interferometer, are not as sensitive to diffuse structures, but if the angular scales of interest are appropriate to the interferometer spacings, an interferometer can be effective.

The MWA (\cite{2013PASA...30....7T}) operates in the frequency range 80 - 300 MHz, with maximum baselines (during the period that describes this work) of 3 km, and with an array configuration that emphasises short baselines and high surface brightness sensitivity.  The MWA shares many physical characteristics with the low frequency SKA, via scaling relations (for example ratio of station diameter to maximum baseline length), and is therefore an excellent instrument with which to make a first exploration of SKA science.  In particular, given the predictions of \cite{Kar:2019mcq}, it is worth exploring DM annihilation scenarios at low frequencies with the MWA, as a precursor study to SKA investigations.  An additional advantage of the MWA for DM studies of dSph galaxies is the survey efficiency, which has led to the ability to report here results for a large sample (relative to prior study sample sizes).

We analyse MWA radio synchrotron data for 14 dSph galaxies, for the first time at frequencies less than 1 GHz. The limits on such synchrotron emission are presented.  We compare these limits to signals predicted from different DM annihilation channels, also considering the future potential of the MWA after recent upgrades.


No new observations or data processing were performed. Data were extracted from existing survey image databases for analysis, specifically the MWA GLEAM survey \cite{2017MNRAS.464.1146H,2015PASA...32...25W} and the TGSS ADR1 \cite{2017A&A...598A..78I}.

Our sample consists of the 14 dSph galaxies from Table 2 of \cite{2014ApJ...783....7C} between declinations $+$30$^{\circ}$ and $-$55$^{\circ}$, being the northern and southern limits of the GLEAM and TGSS ADR1 surveys, respectively.  The sample is listed in Table 1. 

\begin{center}
\begin{table*}
\begin{threeparttable}
\begin{tabular}{c c c c c c c c} 
\hline 
\label{tab1}
TARGET & RA & DEC & Dist. & r$_{half}$ & $S_{RMS}$ & S$_{pred.}$ & Refs.\\ 
 &(hms)&(dms)&(kpc)&(pc)&(mJy/beam)&(mJy/beam) & \\ \hline
Sc1&01h00m09.35s&$-$33d42m32.5s&72&260&14&0.03&\cite{2009ApJ...704.1274W, 1995MNRAS.277.1354I} \\
LeoT&09h34m53.4s&17d03m05s&407&178&27&0.08(0.06)&\cite{2009ApJ...704.1274W, 2008ApJ...684.1075M} \\
LeoIV&11h32m57s&$-$00d32m00s&160&116&32&0.006&\cite{2009ApJ...704.1274W, 2008ApJ...684.1075M} \\
Com&12h26m59s&23d55m09s&44&77&63&0.1&\cite{2009ApJ...704.1274W, 2008ApJ...684.1075M} \\
LeoI&10h08m27.4s&12d18m27s&198&246&32&0.04&\cite{2009ApJ...704.1274W, 1995MNRAS.277.1354I} \\
LeoII&11h13m29.2s&22d09m17s&207&151&35&0.04&\cite{2009ApJ...704.1274W, 1995MNRAS.277.1354I} \\
Car&06h41m36.7s&$-$50d57m58s&85&241&31&0.005&\cite{2009ApJ...704.1274W, 1995MNRAS.277.1354I} \\
For&02h39m59.3s&$-$34d26m57s&120&668&23&0.05&\cite{2009ApJ...704.1274W, 1995MNRAS.277.1354I} \\
Sex&10h13m02.9s&$-$01d36m53s&83&682&20&0.01&\cite{2009ApJ...704.1274W, 1995MNRAS.277.1354I} \\
Boo&14h00m06s&14d30m00s&66&242&47&0.15&\cite{2009ApJ...704.1274W, 2008ApJ...684.1075M} \\
Herc&16h31m02s&02d12m47s&132&330&35&0.001&\cite{2009ApJ...704.1274W, 2008ApJ...684.1075M} \\
LeoV&11h31m09.6s&02d13m12s&180&42&22&0.02&\cite{2009ApJ...704.1274W, 2008ApJ...686L..83B} \\
Seg&10h07m04s&16d04m55s&23&29&30&0.04(0.03)&\cite{2009ApJ...704.1274W, 2008ApJ...684.1075M} \\
Seg2&02h19m16s&20d10m31s&30&34&26&0.05&\cite{2009ApJ...704.1274W, 2009MNRAS.397.1748B} \\ \hline
\end{tabular}
\caption{List of target galaxies}
\begin{tablenotes}
\item Column 1 - Target galaxy name; Column 2 - Right Ascension (hms) of galaxy centroid; Column 3 - Declination (dms) of galaxy centroid; Column 4 - Distance (kpc); Column 5 - Half light radius of galaxy (pc); Column 6 - Measured surface brightness RMS in difference image (mJy/beam); Column 7 - predicted peak surface brightness due to dark matter annihilation (mJy/beam).
\end{tablenotes}
\end{threeparttable}
\end{table*}
\end{center}

The Full width at half maximum (FWHM) of the MWA synthesised beam for the GLEAM survey varies across its 72 - 231 MHz frequency range and in the 170 - 231 MHz band used in this work is typically 2 - 3 arcmin.  Thus, in order to separate point sources from the potential diffuse radio structures of interest, we utilise the TGSS ADR1 survey conducted with the GMRT, at a similar frequency to GLEAM but with an approximate 6 arcsec angular resolution.

For each of the galaxies in Table 1, we downloaded a $5^{\circ}\times5^{\circ}$ image from the GLEAM image server\footnote{http:\/\/mwa-web.icrar.org\/gleam\_postage\/q\/form} and a $1^{\circ}\times1^{\circ}$ image from the TGSS ADR1 image server\footnote{https:\/\/vo.astron.nl\/tgssadr\/q\_fits\/cutout\/form}.

The GLEAM images were regridded to match the TGSS ADR1 images, using the MIRIAD \cite{1995ASPC...77..433S} task regrid.  The TGSS ADR1 images were then convolved with an appropriate Gaussian beam such that the final resolution matched the corresponding GLEAM image, using the MIRIAD task convol.  A scaled version of the convolved TGSS ADR1 image was then subtracted from the regridded GLEAM image, to subtract the point sources detected with TGSS ADR1 from the GLEAM images, using the MIRIAD task maths.

Ideally, this process would produce a difference image that contains only the diffuse emission.  In practise, a range of effects mean that some errors in the difference images are likely.  For example, different ionospheric conditions and applied corrections for the GLEAM and TGSS ADR1 data will cause small mismatches in the positions of point sources, and therefore residual errors in the difference image.  Assuming a single scaling (amounting to a single assumed spectral index) between the GLEAM and TGSS ADR1 images will lead to residual errors in the difference image, due to a range of spectral indices across the population of point sources.

However, we find that generally this process works very well, with very few examples of significant errors.  We are most interested in the difference images in the vicinity of the target galaxies and in these regions we find no significant errors.  In general, across the 14 galaxies, we find noise-like difference images that reflect the confusion-limited signals and diffuse emission expected from the MWA, once point sources are removed.

Figure 1 shows examples of the images and difference images, covering a range of dSph galaxy mass-to-light ratios.  The RMS values measured in each of the 14 difference images are listed in Table 1.  No excess diffuse emission was detected at the locations of the 14 galaxies.

\begin{figure*}
\includegraphics[angle=270,width=0.3\textwidth]{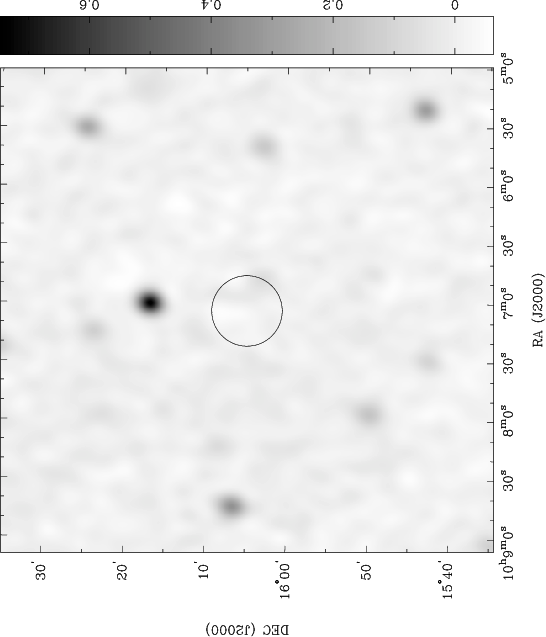}
\includegraphics[angle=270,width=0.3\textwidth]{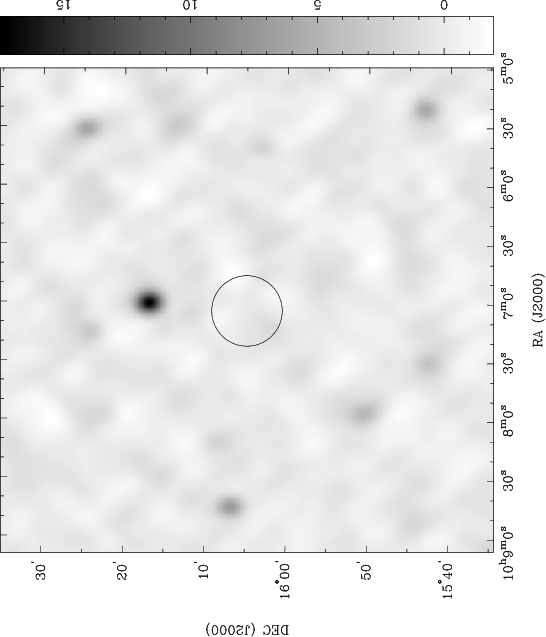}
\includegraphics[angle=270,width=0.3\textwidth]{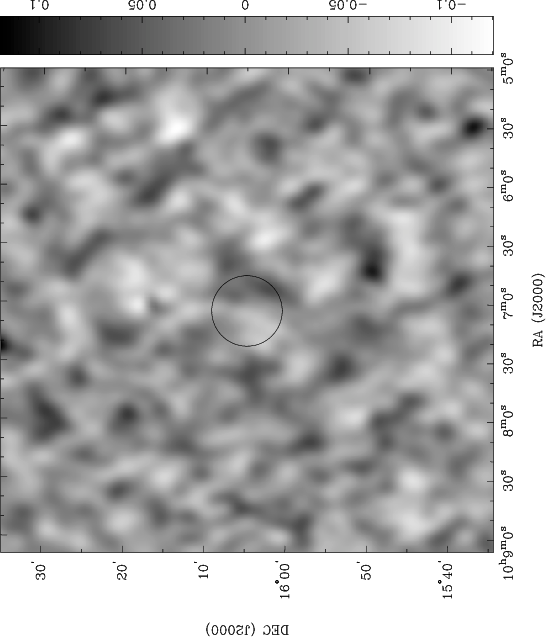}
\includegraphics[angle=270,width=0.3\textwidth]{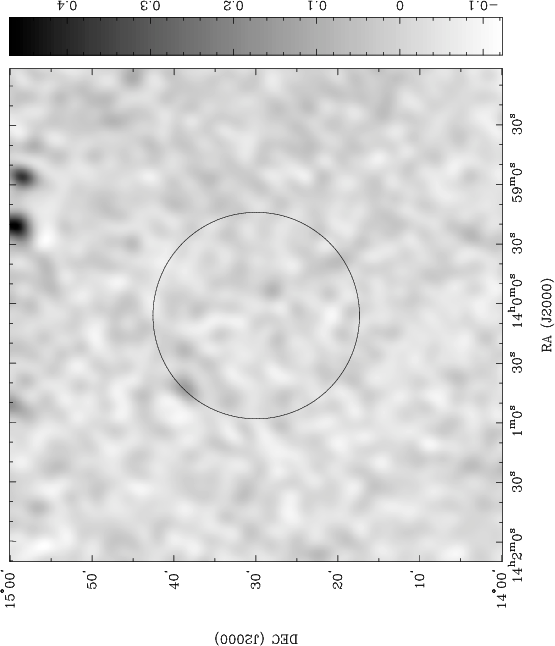}
\includegraphics[angle=270,width=0.3\textwidth]{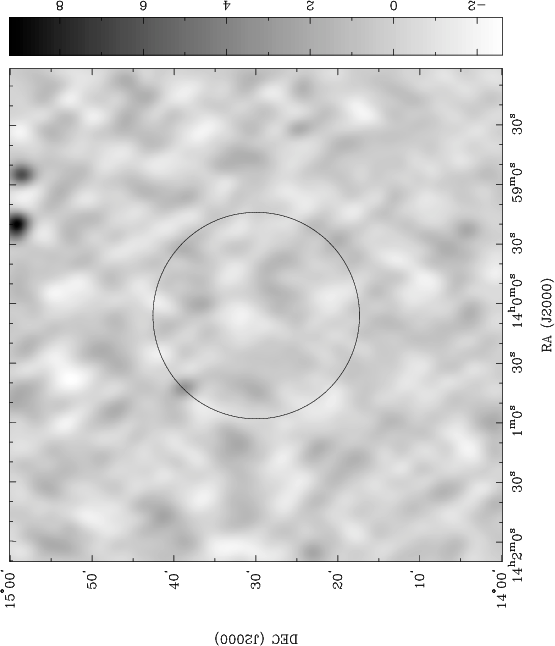}
\includegraphics[angle=270,width=0.3\textwidth]{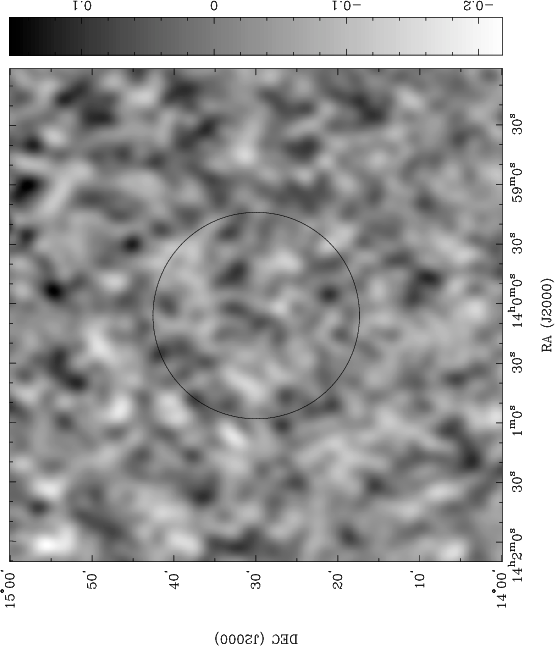}
\includegraphics[angle=270,width=0.3\textwidth]{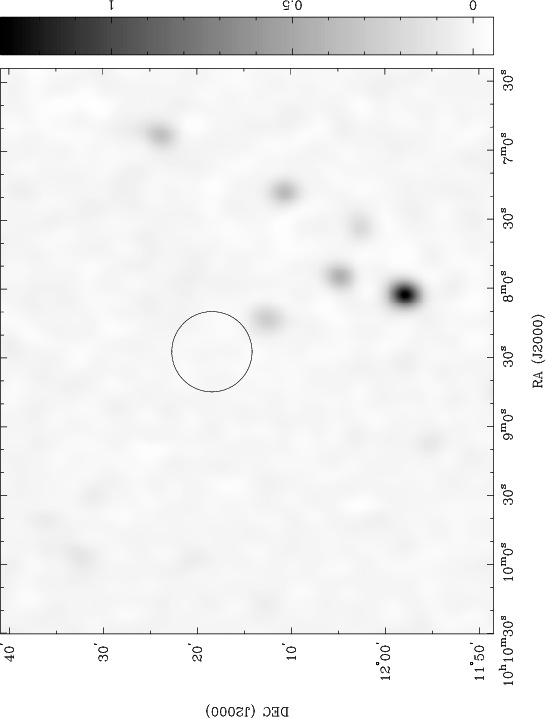}
\includegraphics[angle=270,width=0.3\textwidth]{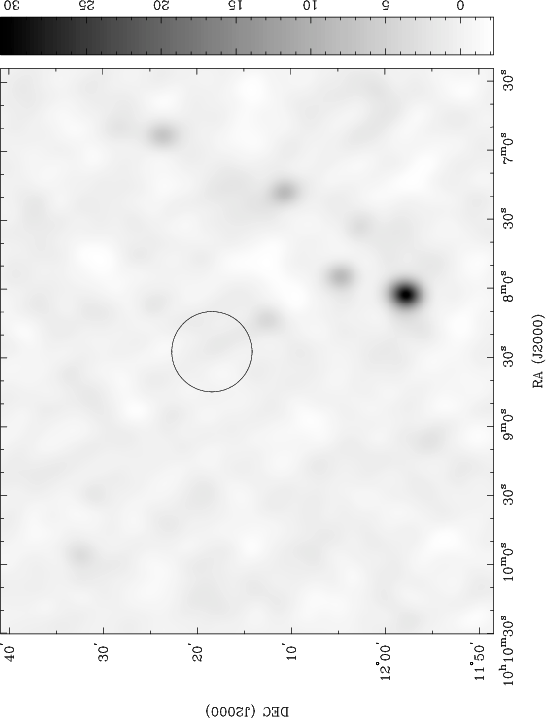}
\includegraphics[angle=270,width=0.3\textwidth]{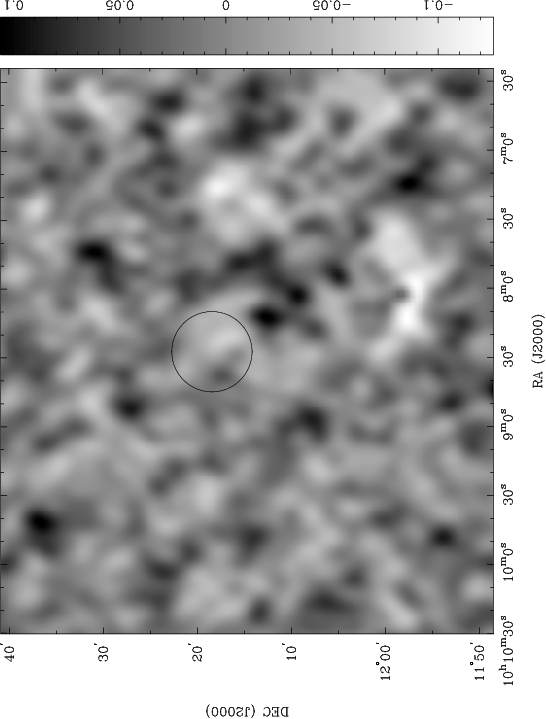}
\caption{GLEAM images (left panels), TGSS ADR1 images convolved to GLEAM resolution (middle panels), the difference images (right panels), for three example target galaxies of varying mass to light ratios \cite{2014ApJ...783....7C}: Segue1 (top: $M/L\sim1400$); Bootes (middle: $M/L\sim200$); and LeoI (bottom: $M/L\sim7$).  The intensity scales for the convolved TGSS ADR1 images are artificially high, as they are not normalised after convolution.  However, the normalisation is absorbed into the scaling applied to match the GLEAM intensity scale.}
\label{fig:fields}
\end{figure*}

For a sample of 14 objects, each being a non-detection, an obvious technique to explore is stacking, whereby the 14 difference images are averaged together to reduce the noise-like contributions.  All of the difference images were regridded such that the galaxy centroid coordinates were centred on the middle pixel of a 512$\times$512 pixel image, with 6 arcsec pixel sizes, using the MIRIAD task regrid.  All 14 centroided difference images were averaged, using the MIRIAD task maths, obtaining an RMS of approximately 9 mJy/beam, with a beam area (defined by the FWHM of the elliptical Gaussian beam) of approximately 4 square arcmin, giving an RMS surface brightness of approximately 2 mJy/arcmin$^{2}$.  No diffuse emission is detected above this level in the stacked image.



The energy distribution of the $e^{\pm}$ originating from DM annihilation in a dSph, which depends on DM mass $m_{\chi}$, the velocity averaged annihilation rate $\langle \sigma v\rangle$ inside the galaxy, and the DM density profile $\rho_{\chi}(r)$, can be obtained for any annihilation channel by using equation (1) of \cite{Kar:2019mcq}. Following equation (2) these $e^{\pm}$ pairs then diffuse and loose energy through the interstellar medium of the galaxy up to large distances and attain a steady state depending on the diffusion parameter ($D(E) = D_0 (E/$1 GeV$)^{0.3}$ (\cite{2007PhRvD..75b3513C, 0004-637X-773-1-61})) and energy loss coefficient ($b(E)$). These charged particles accelerate in the presence of the in-situ magnetic field (B) which leads to the synchrotron radiation \cite[see, e.g.,][]{2017JCAP...09..027M, 2006A&A...455...21C}. The surface brightness expected to be observed by a telescope is obtained by convolving the theoretical signal with the telescope beam.

The nature and properties of particle DM, if it exists, are yet unknown. In view of the consequent lack of knowledge in its annihilation rate in a dSph, the best one can do is to use the available data to constrain the DM parameter space. Such constraints crucially depend on $\rho_{\chi}(r)$, B, and $D_0$. For the dark matter density, we use the NFW profile with $\gamma_{dSph}$ = 1 for Sc1, LeoT, LeoIV, LeoI, LeoII, Car, For, Sex, Herc, and Seg galaxies while for the remaining galaxies, we choose the Einasto profile with $\alpha_{dSph}$ = 0.4 \cite{2018PhRvD..97d3017C}. For the latter class, well-constrained NFW parameters are mostly not available. In the former category, we have checked that NFW and Einasto best-fit parameters \footnote{These best-fit halo parameters are obtained from stellar kinematic data as described in \cite{2018PhRvD..97d3017C}.} lead to fluxes of the same order. The study neglects substructure effects within DM halos, predicted to be small in dSph galaxies (\cite{2013PhRvD..88h3535N, STRIGARI20131}). The radius of the diffusion zone (typically twice the size of the luminous extent of a galaxy) has been set by scaling with respect to either Seg (for smaller galaxies like Com, LeoV, Seg, and Seg2 in Table 1) or Draco (for larger galaxies) using the guidelines discussed in \cite{2015arXiv150703589N,2017JCAP...09..027M,2007PhRvD..75b3513C,0004-637X-773-1-61}.
It is extremely challenging to gain observational insights (say, through
polarization measurements) into the magnetic field properties of dSph galaxies. The lack of any strong observational
lower limits suggests that the magnetic fields could be, in principle,
extremely low. On the other hand, there may be numerous effects that can
give rise to significant magnetic field strengths in dSph. Various theoretical
arguments are proposed for values of $\sim \mu G$ levels.
For detailed discussions we refer the reader to \cite{Regis:2014koa,2011A&A...529A..94C,Regis:2017oet}.
Similarly, little is known about the value of the diffusion coefficient, $D_{0}$, for dSph galaxies; it could be as low as an order of magnitude smaller than that for
the Milky Way \cite{2013PhRvD..88h3535N,2015arXiv150703589N,Jeltema:2008ax}.
Thus, in the absence of any direct knowledge of magnetic field and diffusion
coefficient values for dSph galaxies, we take their values to be $B$ = 1 -- 2 $\mu G$
\footnote{Note that, while Milky Way magnetic field $\sim$ 2 $\mu G$ can be realistic for nearby dwarf galaxies like Seg or Seg2, it might not be the case for other more distant dSphs where this value can be as low as a fraction of $\mu G$ \cite{Regis:2014koa}.}
and $D_0 = 3\times 10^{26} \mbox{cm}^2 \mbox{s}^{-1}$. 
This leads to the largest possible values of flux that one could get from
the current analysis. Stronger magnetic fields and lower values of $D_0$ are disfavoured by already existing observations \cite{2015arXiv150703589N, 2013PhRvD..88h3535N, 2017JCAP...09..027M, 0004-637X-773-1-61,Regis:2014koa}. DSph's which have larger $D_0$ and
smaller magnetic fields (i.e. more conservative choices) would lead to much
lower signals. As will be seen below, our benchmark astrophysical parameters help in probing the maximum allowed range of the DM
parameter space which can be constrained by MWA observations.


Figure 2 represents a model-independent description of DM scenarios
that can be compared to the MWA data presented here (from MWA Phase I observations), as well as those data
expected from its next phase of operations (Phase II operations). Phase II of the MWA contains a new short-baseline array providing even higher surface brightness sensitivity at approximately 15 arcmin angular resolution \cite{2018arXiv180906466W}. The higher surface brightness of MWA Phase II allows the integration of lower surface brightness synchrotron emission to larger radii, meaning that the limits are improved on Phase I in proportion to the change in angular resolution. These Phase II limits have been derived integrating the modeled synchrotron emission over the realistic beam produced from an idealised Phase II observation.  These limits are illustrative only, and observational limits would depend on the exact details of any given observation.

We present results for Boo in Figure 2, for which our predictions are
most encouraging for detection among the 14 dSphs in Table 1.  The figure shows the
minimum $\langle \sigma v\rangle$ corresponding to any $m_\chi$, which will produce radio synchrotron
emission at the RMS thresholds from our Phase I observations, for magnetic fields (B) of 1 $\mu G$ (left panel) and 2 $\mu G$ (right panel).  This is separately estimated for two channels of DM
annihilation, namely, $b\bar{b}$ and $\tau^+ \tau^-$.  The
corresponding plots for the $W^+ W^-$ and $t\bar{t}$ channels fall in
between these two curves.  It may be concluded that any candidate DM
scenario yielding $\langle \sigma v\rangle$ above the curve for MWA Phase
I is excluded by current data, for the choice of astrophysical parameters indicated in the
caption. At the same time, the broken lines indicate the maximum
values of $\langle \sigma v\rangle$ consistent with Fermi-LAT and
cosmic ray data.  Limits from the Phase II MWA exclude more of the parameter space.

\begin{figure*}
\begin{centering}
\includegraphics[angle=0,width=0.45\textwidth]{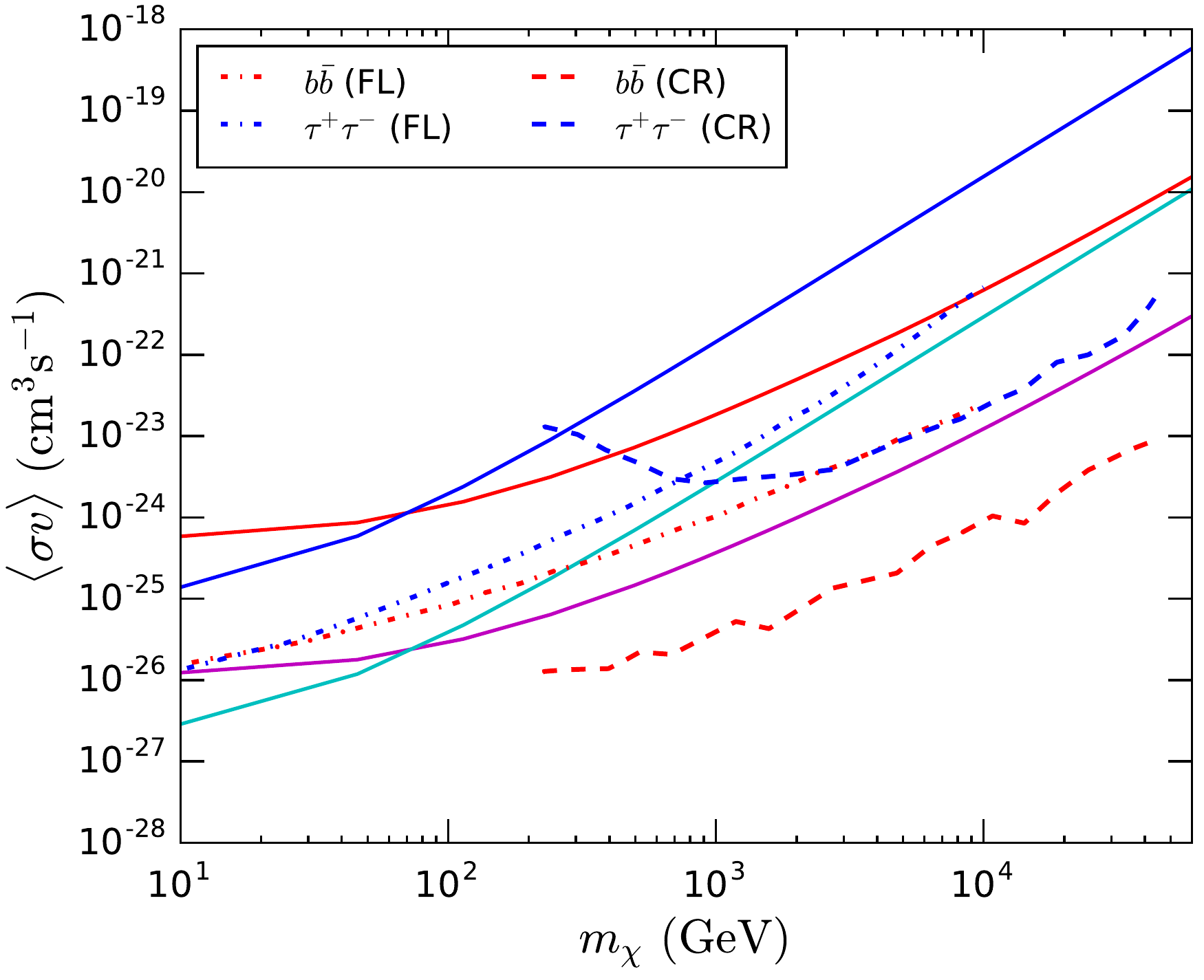}\hspace{10mm}
\includegraphics[angle=0,width=0.45\textwidth]{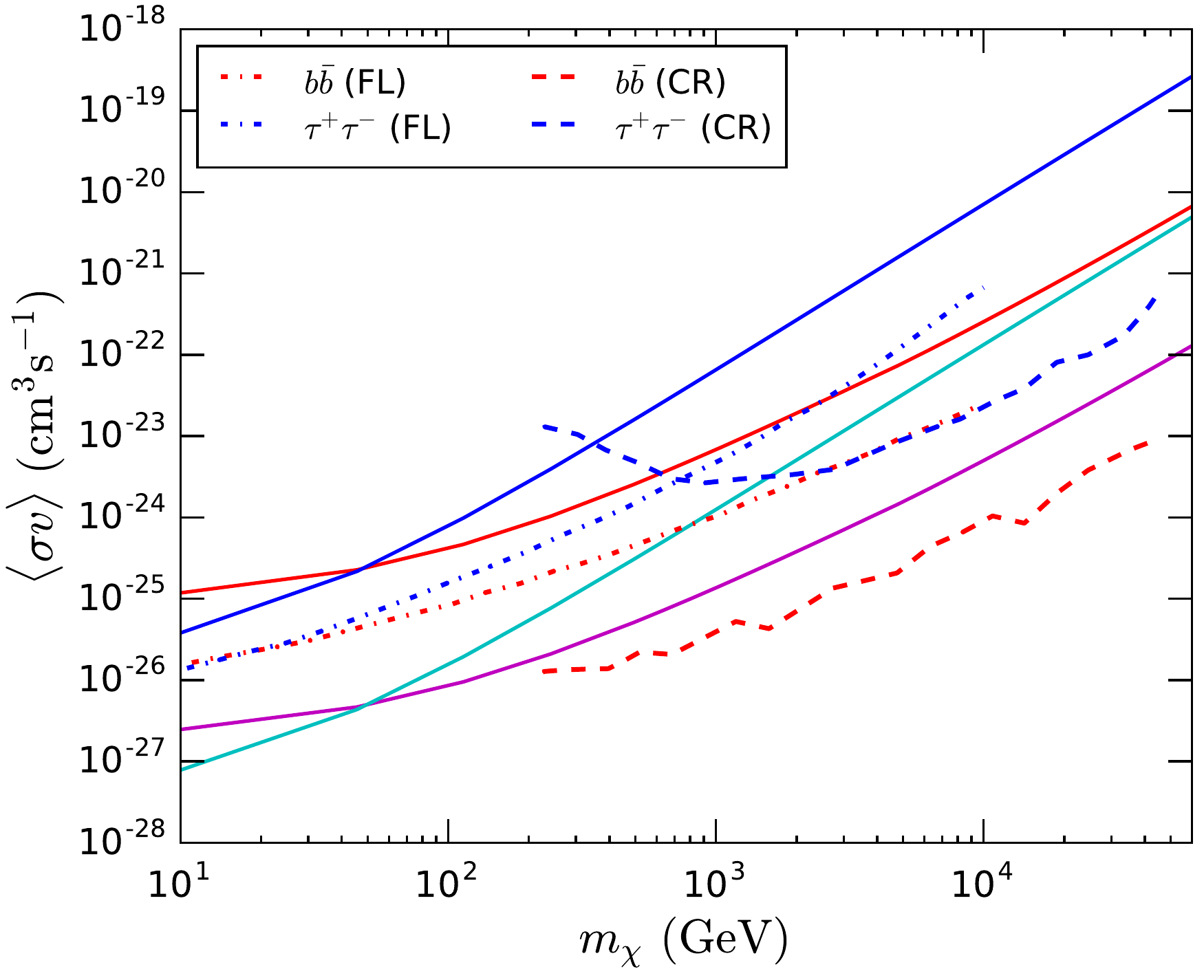}
\includegraphics[angle=0,width=0.45\textwidth]{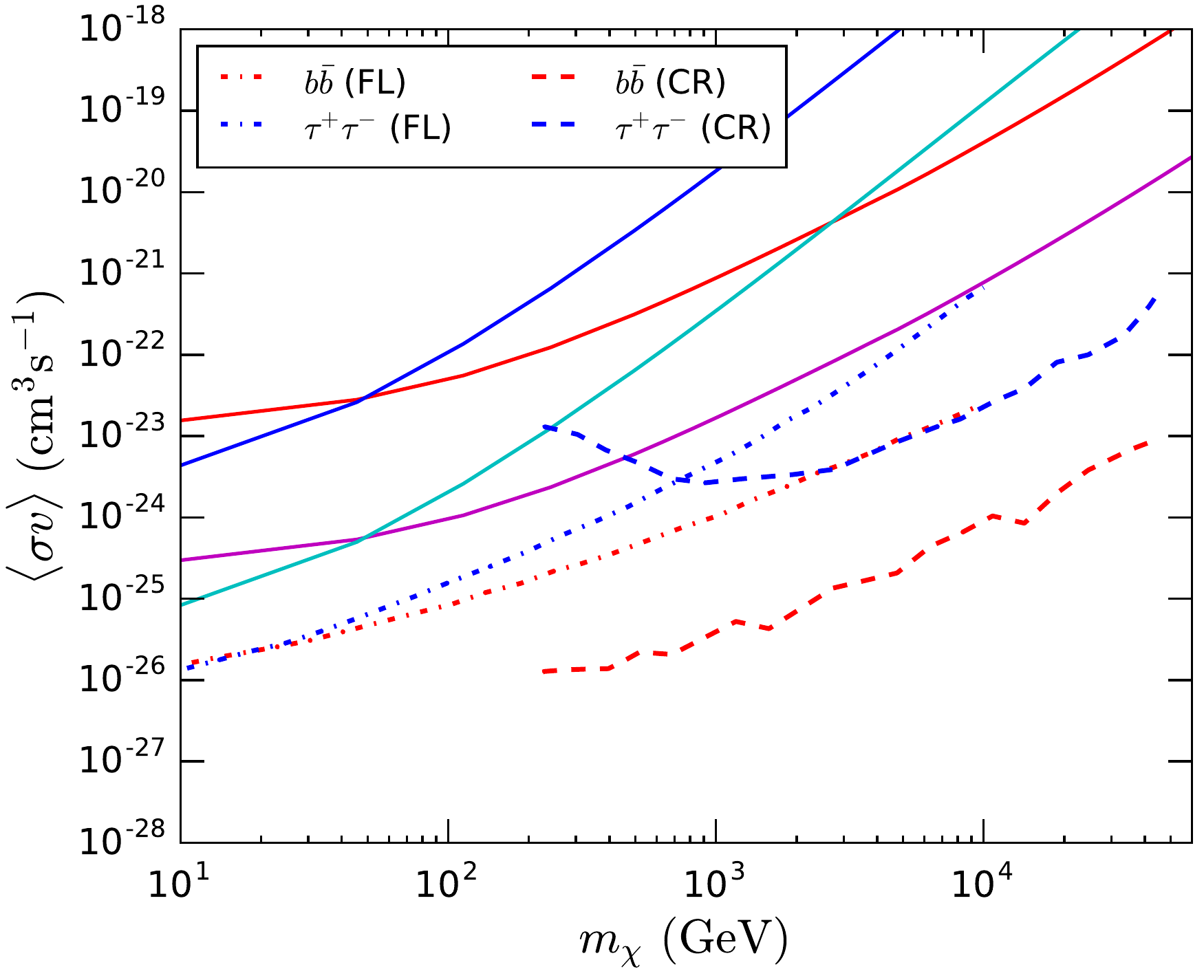}\hspace{10mm}
\includegraphics[angle=0,width=0.45\textwidth]{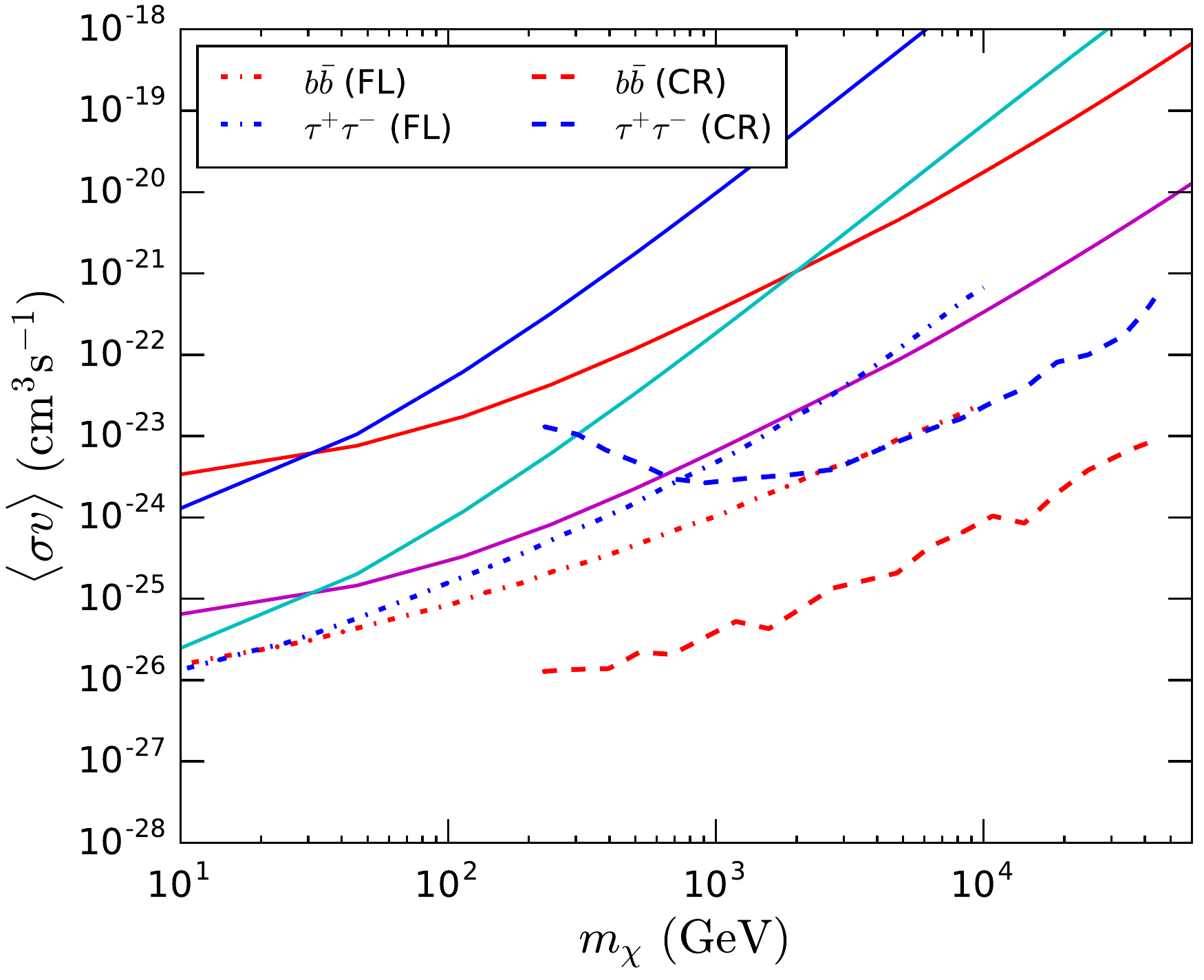}
\caption{{\it Upper panel:} Lower limit (solid lines) in the $\left\langle\sigma v\right\rangle - m_\chi$ plane to observe a signal with the Phase I MWA from Boo galaxy for two different DM annihilation channels, $b\bar{b}$ (solid red) and $\tau^+\tau^-$ (solid blue). The values of the diffusion coefficient and magnetic fields are 
$D_0=3\times10^{26}$ cm$^2$s$^{-1}$ and $B=1$ $\mu$G (left) and $B=2$ $\mu$G (right).
The dashed and dash-dotted lines represent the 95\% C.L. upper limits from cosmic-ray (CR) antiproton observation \cite{2018JCAP...04..004C} and 6 years of Fermi-LAT (FL) gamma-ray data of 15 dSphs \cite{PhysRevLett.115.231301} respectively. The solid magenta ($b\bar{b}$) and solid cyan ($\tau^+\tau^-$) lines show the corresponding limits in Phase II MWA.
{\it Lower panel:} Same as upper panel but with $D_0=3\times10^{28}$ cm$^2$s$^{-1}$.}
\label{fig:stack}
\end{centering}
\end{figure*}

We find that, for B = 1 -- 2 $\mu G$, the predictions for minimum  $\langle \sigma v\rangle$ are already above
the upper limits, even for a non-conservative choice of diffusion 
coefficient, $D_0$ = $3\times 10^{26} \mbox{cm}^2 \mbox{s}^{-1}$. Thus all particle DM scenarios which satisfy the (Fermi-Lat + CR) data are
consistent with the Phase I MWA data. The minimum  $\langle \sigma v\rangle$
lines with Phase II for 1 $\mu G$, on the other hand, are consistent with the CR/Fermi-Lat
limits, for $m_\chi \lesssim$ 200 GeV in the $b\bar{b}$ channel and
 $m_\chi \lesssim$ 1000 GeV (or 1600 GeV for B = 2 $\mu G$) in the  $\tau^+ \tau^-$ channel. 
 We also find that higher values of $D_0$, such as $3\times 10^{28} \mbox{cm}^2 \mbox{s}^{-1}$, which is a rather conservative choice for dSphs considered in this work \cite{2007PhRvD..75b3513C}, can not possibly constrain any DM scenario for both Phase I and Phase II, as shown in the two lower panels of Figure 2.  
 \footnote{However, we have explicitly checked that (not shown in the current paper) for higher magnetic field such as B = 5 $\mu G$, Phase II MWA data can constrain models up to at least $m_\chi \sim$ 500 GeV in the $b\bar{b}$ channel and $m_\chi \sim$ 2500 GeV in the $\tau^+ \tau^-$ channel.
 Higher $D_0$ ($= 3\times 10^{28} \mbox{cm}^2 \mbox{s}^{-1}$) can bring
down these explorable limits of $m_\chi$ to about 200 GeV or 50 GeV respectively. Note that B = 5 $\mu G$ magnetic field is somewhat less realistic for the dSph galaxies considered here \cite{Regis:2014koa}.}

Column 7 of Table 1 shows the predicted peak surface brightness for Phase I MWA due to DM annihilation for all 14 galaxies with minimal supersymmetric standard model (MSSM) benchmark B3 from \cite{Kar:2019mcq}. For some galaxies (for which we have assumed NFW profile), corresponding predictions for Einasto are within brackets. These results are for the choice $D_0 = 3\times 10^{26} \mbox{cm}^2 \mbox{s}^{-1}$ and B = 1 $\mu G$ and the numbers clearly show that the predictions due to this benchmark is always lower than the RMS values for all 14 galaxies.


Detectability at Phase I MWA, in the DM mass range 10 GeV - 50 TeV, requires annihilation cross sections that are already ruled out by gamma-ray and cosmic-ray antiproton observations. Phase II MWA can do significantly better and probe regions still allowed, especially if targeted to sources, such as Boo.  And ultimately the SKA will challenge a very wide range of DM annihilation models.
On the whole, in addition to the exploitation of low-frequency flux, our study improves on existing knowledge in the following way: any positive signal in Phase II will point towards either magnetic field on the higher side ($> 2$ $\mu G$) or a diffusion coefficient at the lower end
($\approx 3\times 10^{26} \mbox{cm}^2 \mbox{s}^{-1}$). An exception can be in the form of $\left\langle\sigma v\right\rangle$ higher than what is predicted in our benchmark \cite{Kar:2019mcq} by about two orders of magnitude, which in tern contradicts the WIMP hypothesis itself.

\section*{Acknowledgments}

This scientific work makes use of the Murchison Radio-astronomy Observatory, operated by CSIRO. We acknowledge the Wajarri Yamatji people as the traditional owners of the Observatory site. Support for the operation of the MWA is provided by the Australian Government (NCRIS), under a contract to Curtin University administered by Astronomy Australia Limited. We acknowledge the Pawsey Supercomputing Centre which is supported by the Western Australian and Australian Governments. AK and BM were partially supported by funding available from the Department of Atomic Energy, Government of India, for the Regional Centre for Accelerator-based Particle Physics (RECAPP), Harish-Chandra Research Institute.

{\it Facility:} Murchison Widefield Array (MWA), Giant Metre-wave Radio Telescope (GMRT)


%

\end{document}